\documentclass[runningheads]{llncs}
\usepackage[T1]{fontenc}
\usepackage{graphicx}
\usepackage{tabularx}
\usepackage{caption}
\usepackage{subcaption}
\usepackage{tabularx}
\usepackage[table,xcdraw]{xcolor}
\usepackage{array}
\usepackage{threeparttable}
\usepackage{url}
\begin{document}
\title{Bridging the Digital Divide: Approach to Documenting Early Computing Artifacts Using Established Standards for Cross-Collection Knowledge Integration Ontology}
\titlerunning{Bridging the Digital Divide: Documenting Early Computing Artifacts}
%
\author{Maciej Grzeszczuk\inst{1,2}\orcidID{0000-0002-9840-3398} \and
Kinga Skorupska\inst{1}\orcidID{0000-0002-9005-0348} \and
Grzegorz Marcin Wójcik\inst{3}\orcidID{0000-0002-4678-9874}}
\authorrunning{M. Grzeszczuk et al.}
%
\institute{Polish-Japanese Academy of Information Technology, Warsaw, Poland \and
The Foundation for the History of Home Computers, Warsaw, Poland
\email{maciej.grzeszczuk@fhkd.pl}  -  
\url{https://fhkd.pl/} \and
University of Maria Curie-Skłodowska, Lublin, Poland}
\maketitle              
\vspace{-0.5cm}
\begin{abstract}
In this paper we address the challenges of documenting early digital artifacts in collections built to offer historical context for future generations. Through insights from active community members (N=20), we examine current archival needs and obstacles. We assess the potential of the CIDOC Conceptual Reference Model (CRM) for categorizing fragmented digital data. Despite its complexity, CIDOC-CRM proves logical, human-readable, and adaptable, enabling archivists to select minimal yet effective building blocks set to empower community-led heritage projects.

\keywords{digital heritage \and software preservation \and historical context \and knowledge systems \and reference model \and museum archives \and ontology}
\end{abstract}

\section{Introduction}

In 1537, King Francis I issued a law ordering publishers in France to deliver a copy of every printed material to the Royal Library, to be archived for future generations. The idea slowly spread to other countries. The UK adopted legal deposit system in the 1662 \cite{roudik2018digitallegal}, while New Zealand did so in the early 20th century \cite{cadavid2017newzealand}.
Today, cultural expressions manifest in various digital formats, including e-books, online articles, multimedia content, video games and social media streams \cite{skold2017expanded}. The adaptation process had started \cite{roudik2018digitallegal}, but there is already nearly 50 years of digital history that needs to be preserved and curated \cite{constantinidis2016crowdsourcing,grzeszczuk2023preserving}.
Inadequate systemic support does not mean, however, that the digital landscape of the end of the 20th century is completely lost. Enthusiasts and independent entities run their own initiatives\footnote{Notable initiatives include: Internet Archive, The Video Game History Foundation, Rhizome, Atari 8-bit Software Preservation Initiative, CSDb: The C-64 Scene Database, demozoo.org, pouet.net, archives.thebbs.org, and textfiles.com..}, trying to protect scraps that are important from their point of view. For some, it is the video games they played on arcades, floppy disks, demoscene music, electronic magazines, for others rare hardware.

\section{The Community of Users and Archivists}

Just as the interests of the founders of these initiatives are diverse, so are the recipients'. We conducted a survey among a sample of 20 enthusiasts interested in using archives that store historical computer artifacts, as well as people actively involved in their preservation. They were recruited through invite-only thematic Facebook and Discord channels, and asked open-ended questions on their and future users' needs from such repositories (RQ1), what are the shortcomings of the existing collections (RQ2), and if they are archiving, what are their documentation practices (RQ3). Responses were collected online or in the form of transcribed telephone interaction directly into Google Forms. They were then subjected to thematic analysis and affinity diagramming. Gathered expectations regarding the content of databases are presented in Table \ref{table_rq1} - now they are mainly looking (68\%) for library information about stored objects. This is followed by searching for the software itself (58\%) and for technical documentation or software manuals (53\%). For some, it is a sentimental journey, a desire to relive childhood experiences (P1, P3, P4, P6, P7, P11, P20), others have a chance to supplement their past experiences now, when access to information is much wider and they can interact more closely with an object they previously only heard about 
(P4, P8, P14, P16, P17, P19, P20), some want to study old design solutions (P19, P12, P10, P9, P8, P5) and share the knowledge with others (P7, P11, P13, P17, P20). 

\vspace{-0.5cm}
\begin{table}[]
\caption{Thematic analysis of respondents’ responses regarding what they think should be stored in databases, what they are currently looking for, and what future generations may be interested in.}
\vspace{2ex}
\scriptsize
\centering
\begin{tabular}{lcc}
\hline
\rowcolor[HTML]{FAFAFA} 
\multicolumn{1}{c}{\cellcolor[HTML]{EFEFEF}\textbf{\begin{tabular}[c]{@{}c@{}}What type of information is sought \\ and should be in the databases\end{tabular}}} & \textbf{Now}                                                                                         & \textbf{\begin{tabular}[c]{@{}c@{}}By future \\ generations\end{tabular}}       \\ \hline
Computer software                                                                                                                                                 & \begin{tabular}[c]{@{}c@{}}P1, P4, P6, P7, P9, P11, \\ P12, P13, P14, P17, P19\end{tabular}          & P7, P11                                                                         \\ \hline
Technical documentation, manuals                                                                                                                                  & \begin{tabular}[c]{@{}c@{}}P4, P5, P7, P8, P9, \\ P10, P11, P12, P15, P16\end{tabular}               & P7, P8, P10                                                                     \\ \hline
Bibliographic information                                                                                                                                         & \begin{tabular}[c]{@{}c@{}}P2, P3, P6, P7, P8, P9, P12, \\ P13, P15, P14, P16, P17, P19\end{tabular} & \begin{tabular}[c]{@{}c@{}}P3, P4, P7, P8, \\ P12, P14, P20\end{tabular}        \\ \hline
Historical context, references                                                                                                                                    & P7, P8, P13, P14, P16                                                                                & \begin{tabular}[c]{@{}c@{}}P1, P2, P3, P6, P7, \\ P8, P9, P12, P17\end{tabular} \\ \hline
Personal stories, historical photos                                                                                                                               & P2, P4, P14, P15, P17, P18                                                                           & \begin{tabular}[c]{@{}c@{}}P2, P4, P7, P9, \\ P15, P17, P18, P20\end{tabular}   \\ \hline
Rarities (source code, prototypes)                                                                                                                                & P4, P6, P13                                                                                          & P10                                                                             \\ \hline
Pictures of the artifacts, physical details                                                                                                                       & \begin{tabular}[c]{@{}c@{}}P4, P7, P9, P11, \\ P13, P15, P16\end{tabular}                            & \begin{tabular}[c]{@{}c@{}}P4, P7, P11, \\ P16, P18\end{tabular}                \\ \hline
Records of experience with artifacts                                                                                                                              & P5, P15                                                                                              & P1, P16                                                                         \\ \hline
How to start, runtime environment                                                                                                                                 & P5, P6, P17                                                                                          & P16                                                                             \\ \hline
\end{tabular}
\begin{tablenotes}
\vspace{1ex}
\item P20 believes that all the information that can be captured are or will be valuable.
\end{tablenotes}
\label{table_rq1}
\vspace{-2ex}
\end{table}

For future generations, who will no longer have access to witnesses of the first digital artifacts, the center of gravity of expected interest migrates towards history: the historical context and cultural and socioeconomic references (47\%), as well as the personal stories of people from this era (42\%).

P16 mentions that while he expects that in the future there will still be emulators that allow you to run a program on a given platform, 
it would be important to convey the entire experience: starting from the style of packaging\footnote{Packaging differed significantly depending on the time and region of distribution.} and ending with the ability to touch the device, feel its weight, or the resistance of the manipulators. For P1, it is crucial to leverage the collected data and artifacts to craft an engaging narrative that brings to life the daily experiences of users within the contexts in which they lived\footnote{Poland during the Iron Curtain is a good example with the lack of immediate access to information or products.}. This, they believe, could be interesting even for a person who is not interested in computing per se. P6 anticipates that most of the collected artifacts may mainly serve as illustrative material for broader narratives, and in rare cases support specialist research — assuming the archive is relatively complete and structured to facilitate such studies (P20, P15, P14).

The wide range of interests of potential repository recipients, being an observation consistent with a broader, international study conducted in 2023 \cite{grzeszczuk2023preserving}, imposes on the needed solution the requirement to store and integrate a large amount of diverse data, with wide possibilities of integrating them with other sources of knowledge, also beyond the domain. 
Considering digital content, we distinguish between those items that were created as digital entities (born-digital) and digital representations of physical objects (scans of books, but also images of magnetic media or laser scans of the buildings) \cite{lor2012ethical,grzeszczuk2023volterra}. For continued and sustainable access to these artifacts it is necessary not only to ensure data integrity but also to describe the runtime environment and necessary hardware \cite{masenya2021metadata}. Finally, layers of relationships that store cultural context, uses and memories need to be applied \cite{melo2023strategy,gonzalez2012charm} so that they retain their complex, nuanced meaning.

In the case of smaller collections, often managed by individuals, metadata, if any, are often limited to library information. \textit{Atarionline.pl}, maintaining large software base collection uses TOSEC file naming scheme\footnote{TOSEC specification: \url{https://www.tosecdev.org/tosec-naming-convention}} as the only form of metadata. \textit{Atarimania.com} has 16 fields of library parameters, such as "genre" or "publisher", plus screenshots and scanned manuals, where available. It also allows community interaction in a form of comments. \textit{Speccy.pl} and \textit{pouet.net} go one step further, adding external reference links to other databases. Searching the collections, however, is limited to the existing few metadata fields.

\subsection{The Challenges of Volunteer Preservation}

We asked our participants\footnote{Out of 20, 11 of them are actively participating in preservation efforts, 1 uses the databases, has the means to aid the effort but does not do it now, 1 did it in the past, 1 rarely engages in such activities, 1 maintains the on-line database, 4 are only the user of existing repositories and 1 is neither a user nor a contributor.} about their documenting practices. 43\% of those who actively archive do not create any documentation in the process. Those who do, however, usually do it in the simplest possible way ("upload to the existing base or their own web site" was indicated by 3 persons, "library information put on the website" by 1 person, "taking a picture" by 1 person, "entry in the XLS sheet" is done by 2 persons, and 1 person "takes notes in the plain text file").
As the key obstacle they indicate lack of time (5 respondents), lack of motivation to do so (5) and lack of tools and structures to do it properly (4). P7 mentions that he lacks a reliable repository for which such activities could be conducted and which would ensure its appropriate long-term exposure and preservation. Such concerns are also echoed by other respondents, who mention the fragmentation of collections coupled with the difficulty of gaining access to the resources (P14, P15, P16, P18) along with the lack of guidance, engagement, skills and tools to motivate potential volunteers (P4, P12, P15, P16, P18). 

\section{The Practice and Theory of Documenting Collections}

Both large national museums \cite{soar_collections} and private collectors face the persistent challenge of an ever-growing backlog in cataloging their collections. Consequently, if an object is processed and some information remains unrecorded, the likelihood of revisiting that object to fill in the gaps is slim. Thus, the initial attempt to capture the context should be as comprehensive as possible - within the frames of domain relevance.
The individuals carrying out these tasks are often hobbyists volunteering their free time. Therefore, the data that needs to be entered should be strictly related to the activity that was performed, with minimal overhead, so as not to discourage the performer and so that it is entered immediately.
Last but not least, the data model should enable easy integration with other data sources.

We decided to consider the CIDOC Conceptual Reference Model (CRM), developed by the CIDOC CRM Special Interest Group. It is a formal ontology for describing the concepts and relationships used in cultural heritage documentation. Since 2006 it has been recognized as an official ISO standard (now ISO 21127:2023) and has multiple successful deployments \cite{mazurek2011applicability,ruymbeke2018implementation,sanfilippo2020ontological}. It allows for broad and flexible context modeling and, unlike competing solution, Cultural Heritage Abstract Reference Model (CHARM), it assumes deployability without the need for a domain-specific extension from a start \cite{gonzalez2012charm}.  

In CIDOC-CRM we capture all the attributes of an object and the relationships between them that are relevant to our domain. For that purpose we instantiate objects of four basic class groups:
\begin{itemize}
\item Material objects (e.g. buildings, documents, objects)
\item Conceptual objects (e.g. literary works, ideas, algorithms, information)
\item Actors (e.g. people, organizations or groups)
\item Events and Activities (e.g. production process, maintenance, transfer)
\end{itemize}
We then connect them with relations (spatial, temporal, relations with actors and context relations). This allows to add concise information to the database about the action or insight, e.g. on a given (day), (the user) (digitized) (the cassette) using (a tape recorder), the (result) is located in a (file) on (the disk).

By assigning a Type (E55 Conceptual Object holding a defined Value) to an object (using the P2 "has type" property), we are able to create multiple dimensions of classification, allowing each object to be contextualized according to various thematic, historical, or functional perspectives.\footnote{For more information see the CIDOC-CRM definition: \url{https://cidoc-crm.org/sites/default/files/cidoc\crm\version\7.3.pdf}}

Integrating many of these types of information creates a vast graph that we can navigate by exploring data in different dimensions. More importantly, the results of such research can be placed in a database in a similar way, with full contextual information about the sources used, creating new knowledge and becoming an explorable object itself.

\section{Application of CIDOC-CRM}

\subsubsection{Initial Processing of new Audio Cassette} 
\begin{figure}
\centering
\includegraphics[width=\linewidth]{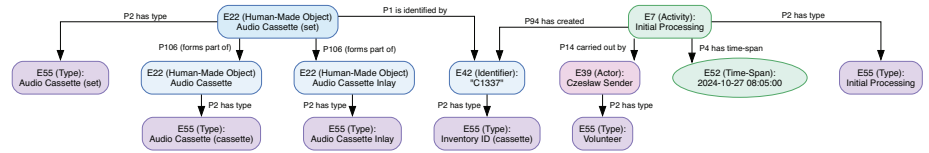}
\caption{CIDOC-CRM v7.3 representation of a cassette tape, forming a set together with its accompanying covers and additional material.}
\label{crm_1}
\end{figure}

We will now consider a case of assigning an Inventory ID to a new cassette\footnote{As performed by volunteers in The Foundation for the History of Home Computers}.
A single cassette usually comes in a plastic box. The box has a paper inlay indicating the content recorded. It can hold additional inserts, such as loading instructions. We want to consider them as a whole, so we model that set as one Human-Made Object (CIDOC-CRM class E22). Assigned Inventory ID is held as a value of E42 Identifier object. We recognize the type of ID held in E42 by the value of E55 Type object that defines it (by a means of P2 "has type" relation between E42 and E55). The types of all the other objects are defined in the same way (see Figure \ref{crm_1}). 

\vspace{-0.3cm}
\vspace{-0.3cm}

According to our experience, tapes are sometimes mistakenly placed in other tapes' storage boxes. Therefore, in our hierarchy, we aim to retain subordinate E22 objects: one for the actual tape as a magnetic media and another for the paper inlay (there may sometimes be more than one). If, in the course of further work, we discover the correct pairings, the data can be reorganized by updating the P106 ("forms part of") assignments. Should multiple tapes arrive from the same source, we want to keep that information. For this case, E78 Curated Holding object will be created, with P106 relations to each of the tape. A source of the tapes can be modeled as E39 Actor, with identifying details such as Name or Address linked in E41, and the relevant E55 applied, to distinguish the type of source. Last but not least, the E7 Action is registered, with E52 Time-Span indication and E39 Actor identifying the individual on the job.

\begin{figure}
\centering
\includegraphics[width=\linewidth]{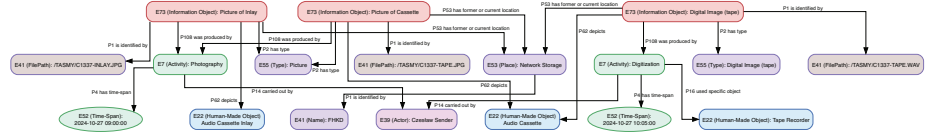}
\caption{CIDOC-CRM v7.3 class structure for products of tape digitization and photo documentation. Redundant objects removed for clarity.}
\label{crm_2}
\vspace{-0.7cm}
\end{figure}

\subsubsection{Documenting the Digitizing Process} Figure \ref{crm_2} illustrates the model we would use for documenting digitization. The goal is to obtain E73 Information Objects, representing files - their path can be found in the assigned E41, and their physical location on the network drive by following P53 "has former or current location" to E53 Place. By observing the P62 "depicts" relation, we can determine what object E73 is in the image, and P16 "used specific object" what tools were used\footnote{We want to keep each of our tape recorders as a separate object, so that we can analyze the quality of the images it generates. We do not have a use for that in the case of a photo camera - those are most often of purely documentary importance and we do not want to clutter the model with information about a specific camera}. Similarly to the previous case, we define and assign the appropriate E55 types, so that the meaning of individual objects is clear to us. For accountability reasons, all operations on a data structure should have an assigned person and indicate the timestamp and scope of the modification via the E7 Activity object and its relationships.

\section{Limitations and Future Work}

The experiment we have presented, at this stage, does not seem to have many advantages. We could effectively encode the same set of information just by establishing a smart file naming scheme \cite{tada2002filenaming} - a common prefix as an Identifier, then a bit for Actor, Tape Recorder, Timestamp, etc. However, the differences will be brought by subsequent structures that we can build on the foundation created here: another Actor can use the inlay photo object, read it and decompose the handwriting into subsequent E73 objects containing a list of titles for each side of the cassette. Each of them can then have a P67 "refers to" relation and point to respective element of another ontology, e.g. describing the game, holding a video of a gameplay or the history of its creation. An Actor, who can decode the contents of the cassette, can connect the decoded binary file to the object containing the raw audio wave. The next Actor can take that binary and try it on a historical computer emulator, checking whether it loads correctly. And if it does, does the loaded game title correspond to the one referenced on the cover? If so, he confirms this fact by creating an appropriate relationship in the model. In this way, step by step, in the form of individual, atomic actions, knowledge is built up that constitutes actual value from using ontology based data integration.

Further work involves expanding the model with additional building blocks to reflect the hierarchy of actions, dependencies and work products from further steps in the process of archiving and preserving cultural heritage.

\section{Discussion and Conclusion}

Although the model uses a relatively large number of small objects to model reality, their function is intuitive and clear to the recipient, even human-readable. 
Due to atomic objects, the amount of data entered should be proportional to the complexity of the task, which in the case of a sequence of repetitive actions on different objects (e.g. in the case of photographing a collection) significantly reduces the overhead. 

Should a CIDOC-CRM extension appear in the future that better reflects the specificity of the work than abstract, high-level models, this does not mean that the work put into collecting and structuring data will go to waste. Thanks to the wide possibilities of integration with other ontologies, the knowledge space created as a result will enable much richer research. New knowledge created as a result of such research can be, while maintaining the entire chain of accountability, stored in the model, potentially increasing its attractiveness. Additionally, although it is not the subject of considerations in this document, organizing data within a standardized structure increases the chance of their survival.

\begin{credits}
\subsubsection{\ackname}The authors express their gratitude to all participants who contributed their time and valuable experiences to this study. We also extend our thanks to the community of enthusiasts who tackle collection, maintenance, and preservation tasks, often without much systemic or institutional support. Leave no tape behind!

\subsubsection{\discintname}
The authors declare that they have no competing interests relevant to this research.
\end{credits}
%
%
%
\bibliographystyle{splncs04}
\bibliography{bibliography}

\end{document}